\documentclass[aps,reprint,prb,longbibliography,superscriptaddress]{revtex4-2}

\usepackage{amsmath}
\usepackage{bbm}
\usepackage{amssymb}
\usepackage{mathtools}
\usepackage{amsthm}
\usepackage{bbm}
\usepackage{amsfonts}
\usepackage{braket}
\usepackage{algorithmic}
\usepackage{lipsum}
\usepackage{tikz}
\usetikzlibrary{positioning,arrows.meta}
\usetikzlibrary{calc}
\usepackage{graphicx}
\usepackage[caption=false]{subfig}  
\usepackage{xcolor}
\usepackage{xfrac}
\usepackage{calc}
\usepackage{accents}

\usepackage{hyperref}

\usepackage{amsthm,letltxmacro,changepage}

\usepackage{tikz}
\usetikzlibrary{shapes.geometric, arrows}
\usetikzlibrary{patterns}
\tikzstyle{startstop} = [rectangle, rounded corners, minimum width=3cm, minimum height=1cm,text centered, draw=black, fill=red!30]

\tikzstyle{box} = [rectangle, rounded corners, minimum width=4cm, minimum height=7cm,text centered, draw=black, fill=purple!30,opacity=0.4]

\tikzstyle{io} = [trapezium, trapezium left angle=70, trapezium right angle=110, minimum width=1cm, minimum height=1cm, text centered, draw=black, fill=blue!30]

\tikzstyle{process} = [rectangle, minimum width=3cm, minimum height=1cm, text centered, draw=black, fill=blue!30,pattern=horizontal lines light blue]
\tikzstyle{decision} = [diamond, minimum width=3cm, minimum height=1cm, text centered, draw=black, fill=green!30]
\tikzstyle{arrow} = [ultra thick,->,>=stealth]

\newcommand{\cfig}[1]{Fig.~\ref{#1}}
\newcommand{\ceqn}[1]{Eq.~\eqref{#1}}
\newcommand{\capp}[1]{Appendix~\ref{#1}}
\newcommand{\cref}[1]{Ref.~\cite{#1}}

\newcommand{\cact}[1]{\hyperref[#1]{#1}}
\newcommand{\ignore}[1]{}

\DeclareUnicodeCharacter{0301}{*************************************}

\DeclareMathOperator{\imag}{\mathfrak{Im}}

\DeclareMathOperator{\zubaL}{\langle\!\langle}
\DeclareMathOperator{\zubaR}{\rangle\!\rangle}
\newcommand{\zuba}[1]{\zubaL #1 \zubaR}

\newcommand{\mat}[1]{\mathbb{#1}} 
\newcommand{\vek}[1]{\mathbf{#1}} 

\def\getfirst#1.#2\relax{#1}

\newcommand{\rwkv}{RWKV}

\definecolor{commcol}{RGB}{0,0,0}
\definecolor{new_text}{RGB}{0,0,0}

\begin{document}


\title{Operator Lanczos Approach enabling Neural Quantum States as Real-Frequency Impurity Solvers}

\author{Jonas B. Rigo}
\email[]{j.rigo@fz-juelich.de}
\affiliation{Forschungszentrum J\"{u}lich GmbH, Peter Gr\"{u}nberg Institute,
	Quantum Control, 52425 J\"{u}lich, Germany}
\affiliation{University of Regensburg}
\author{Markus Schmitt}
\email[]{markus.schmitt@ur.de}
\affiliation{Forschungszentrum J\"{u}lich GmbH, Peter Gr\"{u}nberg Institute,
	Quantum Control, 52425 J\"{u}lich, Germany}
\affiliation{University of Regensburg}

\begin{abstract}
To understand the intricate exchange between electrons of different bands in strongly correlated materials, it is essential to treat multi-orbital models accurately.
For this purpose, dynamical mean-field theory (DMFT) provides an established framework, whose scope crucially hinges on the availability of efficient quantum impurity solvers.
Here we present a real-frequency impurity solver based on neural quantum states (NQS) combined with an operator-Lanczos construction. NQS are an asymptotically unbiased variational ground-state ansatz that employs neural networks to capture long-range correlations on complicated graph structures. We leverage this ability to solve multi-orbital impurity problems using a systematically improvable Segmented Commutator Operator-Lanczos (SCOL) construction. Our benchmarks on both the single-orbital Anderson model and the multi-orbital Hubbard–Kanamori impurity Hamiltonian reveal excellent ground-state precision and the capacity to accurately resolve zero temperature spectral functions and self-energies. These results open avenues for extending DMFT to more challenging problems.
\end{abstract}

\maketitle

Quantum materials exhibit a plethora of exotic quantum phases, often driven by strong correlations that defy descriptions based on non-interacting particles. The intricate interplay of spin, charge, orbital, and lattice degrees of freedom gives rise to rich phase diagrams that are interesting from a technological and scientific perspective \cite{pham20222d,awschalom2013quantum,wolf2001spintronics}. Examples range from Mott insulators in transition metal oxides \cite{park2012site,catalano2018rare} and heavy fermion metals in rare-earth intermetallics \cite{stewart2001non,gegenwart2008quantum}, to unconventional superconductors \cite{keimer2015quantum,cao2018unconventional} and quantum spin liquids in frustrated magnets \cite{zhou2017quantum}. While experimental techniques grant tremendous insights into the phenomenology of quantum materials, their complexity makes it challenging to supplement findings with analytical approaches. 
Therefore, numerical methods are essential for our theoretical understanding.

One powerful framework for studying correlated quantum matter is the dynamical mean field theory (DMFT) \cite{georges1996dynamical,kotliar2006electronic}. Its effectiveness is due to many lattices of physical interest having large coordination numbers, which suppresses nonlocal correlations. DMFT exploits this by mapping the full lattice problem onto a single interacting site coupled to an effective fermionic bath that captures the charge and spin exchange with the rest of the system. This construction is controlled by the assumption of a local self-energy, making the method exact in the limit of infinite coordination number \cite{georges1996dynamical} and remarkably accurate for high-coordination lattices encountered in real materials. 

Even though impurity models are nowadays routinely solved by established methods in standard DMFT, they become significantly more challenging for systems with low coordination numbers or several interacting bands. These cases are treated with extensions of DMFT \cite{maier2005quantum,toschi2007dynamical,rohringer2018diagrammatic} that, however, challenge even state-of-the-art quantum impurity solvers  \cite{rubtsov2005continuous,bauernfeind2017fork,lee2021computing}. In fact, there is no limit to the complexity of the quantum impurity models within DMFT extensions, creating a constant demand for ever more powerful impurity solvers. In this work we address this demand with a new flexible and scalable real-frequency impurity solver on the basis of \textit{neural quantum states} (NQS) \cite{carleo2017solving}.

Although quantum impurity models involve only a small set of interacting degrees of freedom, their irregular connectivity and fermionic statistics make them difficult to solve. Quantum Monte Carlo methods encounter the fermionic sign problem for generic impurity geometries \cite{troyer2005computational,foulkes2001quantum}, which severely limits their applicability, and the widely adopted \textit{continuous-time} quantum Monte Carlo methods only provide reliable access to imaginary-frequency Green's functions \cite{rubtsov2005continuous}. A complementary approach is offered by tensor networks (TN), which can be propagated in real time \cite{bauernfeindForkTensorProductStates2017}, along a complex contour \cite{grundnerComplexTimeEvolution2024} or even employed in a Krylov procedure \cite{gleisProjectorFormalismKept2022a,kovalskaTangentSpaceKrylov2025} to obtain real-frequency propagators. However, intricate entanglement structures resulting from high coordination numbers can limit the reach of TN methods. Neural quantum states (NQS) offer a promising alternative \cite{carleo2017solving}. They provide flexible ans\"atze capable of representing highly entangled wave functions on intricate interaction graphs \cite{sharir2022neural,levineQuantumEntanglementDeep2019a,gaoEfficientRepresentationQuantum2017,dengQuantumEntanglementNeural2017,denisCommentCanNeural2025}, making them naturally suited for the complex topologies characteristic of impurity Hamiltonians. Furthermore, the variational Monte Carlo framework has \textit{no sign problem related to probabilities} \cite{becca2017quantum} and at the same time scales favorably with computational resources. This advantage has enabled fermionic NQS architectures to achieve state-of-the-art performance on challenging systems of correlated electrons \cite{robledo2022fermionic,humeniuk2023autoregressive,guSolvingHubbardModel2025a,rothSuperconductivityTwodimensionalHubbard2025,ibarra2025autoregressive,lange2025tJ,Sobral2024PhysicsinformedTF}. Recent progress on representing the ground state of an impurity model with NQS \cite{caoVisionTransformerNeural2024} underscores their promise as real-frequency impurity solvers.

In the following, we exploit these strengths to construct a real-frequency impurity solver. We introduce the \textit{Segmented Commutator Operator-Lanczos} (SCOL) scheme and combine it with an autoregressive NQS ground state to obtain the impurity Green's function and self-energy directly at zero temperature. We demonstrate the approach on the single-impurity Anderson model (SIAM) and on two- and three-orbital Hubbard-Kanamori Hamiltonians, which constitute a challenging benchmark \cite{kugler2019orbital,grundnerComplexTimeEvolution2024}.

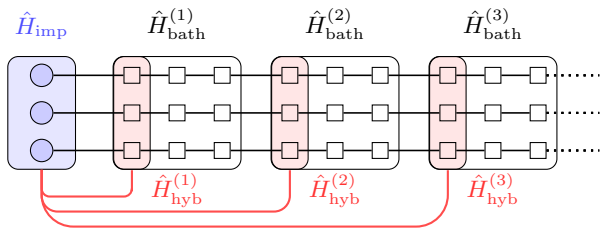
\begin{figure}[t]
    \centering
\begin{tikzpicture}[
    siteimp/.style={circle,draw,fill=blue!20,inner sep=3pt},
    sitebath/.style={rectangle,draw,inner sep=3pt},
    blockbath/.style={rounded corners,draw,inner sep=6pt},
    hybline/.style={thick,red!70},
    >=stealth
]

\node[draw,rounded corners,fill=red!10,minimum width=0.485cm,minimum height=1.5cm] (hybblock) at (1.2,0.3) {};
\node[draw,rounded corners,fill=red!10,minimum width=0.485cm,minimum height=1.5cm] (hybblock) at (3.3,0.3) {};
\node[draw,rounded corners,fill=red!10,minimum width=0.485cm,minimum height=1.5cm] (hybblock) at (5.4,0.3) {};

\node[draw,rounded corners,fill=blue!10,minimum width=0.9cm,minimum height=1.5cm] (impblock) at (0,0.3) {};
\node[siteimp] (imp1) at (0,0.8) {};
\node[siteimp] (imp2) at (0,0.3) {};
\node[siteimp] (imp3) at (0,-0.2) {};
\node[above=3pt of impblock] {\color{blue!70}$\hat H_{\mathrm{imp}}$};

\def\nb{3}       
\def\nrow{3}     

\newcommand{\bathblock}[2]{%
  \begin{scope}[shift={(#2,0)}]
    \node[blockbath,minimum width=1.7cm,minimum height=1.5cm] (bb#1) at (-0.2,0.3) {};
    \foreach \row in {0,...,2} {
      \foreach \col in {0,...,2} {
        \pgfmathsetmacro\x{-0.8 + 0.6*\col}
        \pgfmathsetmacro\y{-0.2 + 0.5*\row}
        \node[sitebath] (b#1-\row-\col) at (\x,\y) {};
      }
    }
    \node[above=3pt of bb#1] {$\hat H_{\mathrm{bath}}^{(#1)}$};
  \end{scope}
}

\bathblock{1}{2.}
\bathblock{2}{4.1}
\bathblock{3}{6.2}

\foreach \col in {0,1}{
\foreach \row in {0,...,2} {
  \pgfmathtruncatemacro{\colp}{\col+1}
  \draw[semithick] (b1-\row-\col.east) -- (b1-\row-\colp.west);
  \draw[semithick] (b2-\row-\col.east) -- (b2-\row-\colp.west);
  \draw[semithick] (b3-\row-\col.east) -- (b3-\row-\colp.west);
  \draw[thick, dotted] (b3-\row-2.east) -- ++(0.7,0);
}
}

\foreach \row in {0,...,2} {
  \draw[semithick] (b1-\row-2.east) -- (b2-\row-0.west);
  \draw[semithick] (b2-\row-2.east) -- (b3-\row-0.west);
  \draw[thick, dotted] (b3-\row-2.east) -- ++(0.7,0);
}
\draw[semithick] (imp3.east) -- (b1-0-0.west);
\draw[semithick] (imp2.east) -- (b1-1-0.west);
\draw[semithick] (imp1.east) -- (b1-2-0.west);

\coordinate (hybBend1) at ($(b1-0-0.south)+(0.0,-0.5)$);
\draw[hybline, rounded corners=3pt]
  (impblock.south) |- (hybBend1) -| ($(b1-0-0.south)+(0.0,-0.15)$);
\coordinate (hybBend2) at ($(b1-0-1.south)+(0.0,-0.7)$);
\draw[hybline, rounded corners=6pt]
  (impblock.south) |- (hybBend2) -| ($(b2-0-0.south)+(0.0,-0.15)$);
\coordinate (hybBend3) at ($(b1-0-1.south)+(0.0,-0.9)$);
\draw[hybline, rounded corners=12pt]
  (impblock.south) |- (hybBend3) -| ($(b3-0-0.south)+(0.0,-0.15)$);
  
\node[below=2pt of b1-0-1.south, text=red!80] {$\hat H_{\mathrm{hyb}}^{(1)}$};
\node[below=2pt of b2-0-1.south, text=red!80] {$\hat H_{\mathrm{hyb}}^{(2)}$};
\node[below=2pt of b3-0-1.south, text=red!80] {$\hat H_{\mathrm{hyb}}^{(3)}$};



\end{tikzpicture}
\caption{Graphical representation of the SCOL decomposition \ceqn{eq:scol_decomp} of a three-orbital Hamiltonian with segment depth \(l_s = 3\) and \(N_s = L_c/l_s\).}
\vspace{-0.5cm}
    \label{fig:scol}
\end{figure}


\textit{Operator-Lanczos}---After obtaining an accurate ground state \(\ket{\psi_0}\) of a many-body Hamiltonian \(\hat{H}\), dynamical correlation functions can be computed by projecting \(\hat{H}\) onto a suitably chosen low-dimensional subspace that captures its dominant excitations. In the traditional \textit{Lanczos procedure} \cite{dargelLanczosAlgorithmMatrix2012}, the subspace is generated by repeatedly applying \(\hat{H}\) in a recursion to generate excited states. However, for many numerical techniques, evaluating expectation values in the ground state is far more efficient and numerically stable than generating intermediate states like \(\ket{\psi_n} = \hat{H}\ket{\psi_{n-1}}\). Therefore we formulate the Lanczos projection directly in operator space by expressing excited states as
\(\ket{\phi_a} = \hat{O}_a \ket{\psi_0}\), where the operators \(\{\hat{O}_a\}\) are the \textit{Lanczos operators} that generate the relevant excitations from the
ground state. 

For a Green's function (GF) \(\mathcal{G}_{AB}(\omega) = \zubaL{}\hat{A};\hat{B}\zubaR{}_{\omega^+}\), the Lanczos operator set \(\{\hat{O}_n\}\) is generated from the \emph{seed operator} \(\hat{O}_1 = \hat{B}\) through the  recursion
\begin{align}
\label{eq:op_lanczos}
\hat{H}\hat{O}_{n-1}\ket{\psi_0}
&=  \hat{O}_{\,n-1}\hat{H}\ket{\psi_0}
+ [\hat{H}, \hat{O}_{n-1}]\ket{\psi_0},
\nonumber\\
&\equiv  E_0\,\hat{O}_{\,n-1}\ket{\psi_0}
+ \hat{O}_n\ket{\psi_0},
\end{align}
with the new operator \(\hat{O}_n = [\hat{H}, \hat{O}_{n-1}]\). To compute \(\mathcal{G}_{AB}\) the left seed operator must also enter the Lanczos operator set \(\hat{O}_0=\hat{A}\). With the completed set one can compute the projected Hamiltonian and \textit{Gram} (or \textit{overlap}) matrices
\begin{equation}
  \label{eq:scol_mats}
  \mat{H}_{ab}
  = \bra{\psi_0}\hat{O}_a^\dagger \hat{H}\hat{O}_b\ket{\psi_0}\ ,
  \quad
  \mat{G}_{ab}
  = \bra{\psi_0}\hat{O}_a^\dagger \hat{O}_b\ket{\psi_0} ,
\end{equation}
which yield the generalized eigenvalue problem
\begin{equation}
\label{eq:gen_eigp}
  \mat{H}\Psi = E\,\mat{G}\Psi .
\end{equation}
The solution of \ceqn{eq:gen_eigp} gives the poles defining the GF following the prescription in \capp{sec:gfcalc}.

A central property of the operator formulation is its \emph{invariance under linear transformations}. Any (invertible) linear combination of the Lanczos operators produces the same spectrum since only the number of distinct operator directions matters, not their specific linear representation (see \capp{sec:invar_proof} for the proof).
For non-interacting Hamiltonians this means that the single-particle excitation space is exactly represented by \(\hat{O}_i=\hat{c}_i^\dagger\), and the set \(\{\hat{c}_i^\dagger\}\) already yields the exact solution for all propagators of the type \(\zubaL{}\hat{c}_i;\hat{c}_j^\dagger\zubaR{}_{\omega^+}\).

\textit{Segmented commutator operator-Lanczos}---%
Quantum impurity Hamiltonians \(\hat H=\hat{H}_{\mathrm{imp}}+\hat{H}_{\mathrm{bath}}+\hat{H}_{\mathrm{hyb}}\) consist of a small interacting system \(\hat{H}_{\mathrm{imp}}\) (the \textit{impurity}) and one or several (\(n_b\)) infinitely large, non-interacting, fermionic \textit{baths} summarized in \(\hat{H}_{\mathrm{bath}}\). The two systems are coupled via a hybridization term \(\hat{H}_{\mathrm{hyb}}\), which can be assumed to only act on the first site of each bath, since non-interacting systems can always be mapped to a semi-infinite chain \cite{wilson1975renormalization}, as shown in \cfig{fig:scol}. Even though the baths get truncated and thus, are in practice finite, the na\"ive recursion of \ceqn{eq:op_lanczos} leads to a prohibitive exponential proliferation of increasingly more complex terms. To control this growth, while retaining contributions from the entire bath, we introduce the \emph{segmented commutator operator-Lanczos} (SCOL) scheme. SCOL controls the term proliferation by dividing the combination of all bath chains into \(N_s\) non-overlapping \emph{segments}. Each segment with index \(\alpha=1,\dots,N_s\) is treated as \(n_b\) independent tight-binding chains connected to the impurity, defining auxiliary Hamiltonians \(1\le \alpha \le N_s\)
\begin{equation}
\label{eq:scol_decomp}
\hat{H}^{(\alpha)} =
\hat{H}_{\mathrm{imp}}
+\hat{H}_{\mathrm{bath}}^{(\alpha)}
+\hat{H}_{\mathrm{hyb}}^{(\alpha)} .
\end{equation}
Within each segment we generate local commutator chains up to \(l_s\) recursions starting from the seed operator \(\hat{O}_0 = \hat{B}\)
(e.g.\ \(\hat{O}_0=\hat{c}_j^\dagger\))
\begin{equation}
\label{eq:comm_rec}
\hat{O}_n^{(\alpha)} = [\hat{H}^{(\alpha)}, \hat{O}_{n-1}^{(\alpha)}],
\qquad n \le l_s .
\end{equation}
During this recursive construction, any term composing \(\hat{O}_n^{(\alpha)}\) that contains more than \(m_{\max}\) fermionic creation or annihilation operators is removed. To get the SCOL set \(\mathcal{B}\), each \(\hat{O}_n^{(\alpha)}\) is decomposed into its constituent normal-ordered operator strings \(\hat{O}_n^{(\alpha)} = \sum_\ell C_{n\ell}^{(\alpha)}\hat{S}_{n\ell}^{(\alpha)}\) and the distinct strings \(\hat{O}_{(\alpha n \ell)} = \hat{S}_{n\ell}^{(\alpha)}\) then form the set \(\mathcal{B} = \{\hat O_{a}\}\) with the index \(a = (\alpha n \ell)\). This decomposition leaves the variational subspace unchanged, but simplifies the estimation in the Monte Carlo scheme discussed below. Finally, to guarantee the correct mean-field limit, we include the bare creation (or annihilation) operators \(\{\hat{c}_i^\dagger\}\) in \(\mathcal{B}\). For any impurity model, the SCOL set is fully defined by \(N_s\) and \(l_s\) which can be varied independently.

SCOL is agnostic to the ground-state solver and can be combined with \textit{tensor network} methods or even \textit{determinant quantum Monte Carlo}, provided that the simulation temperature is below the spectral gap \cite{huangDeterminantalQuantumMonte2022,zhangQuantumMonteCarlo2003}. Within variational Monte Carlo (and thus NQS), it becomes particularly efficient because the matrix elements of \(\mat{G}\) and \(\mat{H}\) can be written as highly parallelizable Monte Carlo expectation values over occupation configurations \(\vek{s}\).

Each projected matrix element \(\mat{G}_{ab}\)
is evaluated as a Monte Carlo expectation value \(\mat{G}_{ab} = \sum_\vek{s} \tfrac{p_{ab}(\vek{s})}{p_{ab}(\vek{s})}\langle\psi_{\pmb{\theta}}|\hat{O}_a^\dagger\ket{\vek{s}}\bra{\vek{s}} \hat{O}_b|\psi_{\pmb{\theta}}\rangle\)  over configurations in the \emph{shifted} quantum-number sector on which \(\hat{O}_a\ket{\psi_{\pmb{\theta}}}\) and \(\hat{O}_b\ket{\psi_{\pmb{\theta}}}\) have support, but not \(|\psi_{\pmb{\theta}}(\vek{s})|^2\), which is otherwise the natural choice for \(p_{ab}(\vek{s})\).
Therefore, for each operator \(\hat{O}_a\) we define an auxiliary distribution \(p_a(\vek{s}) \propto |\langle\vek{s}|\hat{O}_a|\psi_{\pmb{\theta}}\rangle|^2\) that does have the right support. For every pair \((a,b)\) we draw \(N_{ab}\) samples from both \(p_a\) and \(p_b\), construct two independent Monte Carlo estimates of \(\mat{G}_{ab}\) and then select the lower-variance one. Denoting the corresponding distribution by \(p_{ab}\), the estimator is
\begin{equation}
\label{eq:ONestimation}
  \mat{G}_{ab}
  = \mathbb{E}_{\vek{s}\sim p_{ab}(\vek{s})}
    \big[
      \big(L_a(\vek{s})\big)^\dagger
      L_b(\vek{s})
    \big],
\end{equation}
with local estimators
\(L_a(\vek{s}) = \langle\vek{s}|\hat{O}_a|\psi_{\pmb{\theta}}\rangle / \sqrt{p_{ab}(\vek{s})}\).
For a given sampled configuration \(\vek{s}\), all \(L_a(\vek{s})\) can be computed completely independently for all \(\hat{O}_a\in\mathcal{B}\), so the evaluation of \(\mat{G}_{ab}\) is highly parallelizable across both samples and operator indices.

Finally, the projected Hamiltonian matrix elements can be evaluated from the same samples using the commutator identity
\begin{equation*}
  \mat{H}_{ab}
  = \bra{\psi_{\pmb{\theta}}} \hat{O}_a \hat{H} \hat{O}_b\ket{ \psi_{\pmb{\theta}}} 
  = E_0\,\mat{G}_{ab}
  +  \bra{\psi_{\pmb{\theta}}} \hat{O}_a [\hat{H}, \hat{O}_b] \ket{\psi_{\pmb{\theta}}},
\end{equation*}
where \(E_0=\langle \psi_{\pmb{\theta}}|\hat{H}|\psi_{\pmb{\theta}}\rangle\) is the ground-state energy. For fixed \(b\), we introduce the commutator operator \(\hat{O}_\beta = [\hat{H},\hat{O}_b]\) and its local estimator
\(L_\beta(\vek{s})\) defined in complete analogy to \(L_a(\vek{s})\). The mixed term \(\bra{\psi_{\pmb{\theta}}} \hat{O}_a [\hat{H},\hat{O}_b] \ket{\psi_{\pmb{\theta}}}\) is then obtained as an expectation value of \(\big(L_a(\vek{s})\big)^\dagger L_\beta(\vek{s})\) over the \emph{same} configurations \(\vek{s}\) already used for \(\mat{G}_{ab}\), so that only \(L_\beta\) has to be computed in addition. This strategy avoids the explicit and costly application of \(\hat{H}\) to all basis states \(\hat{O}_a\ket{\psi_{\pmb{\theta}}}\), under the mild assumption that the variational state is close to an eigenstate.

\textit{Neural Quantum States}---%
NQS are a compression technique for many-body wave functions \cite{carleo2017solving}.  
The underlying idea is to exploit the strengths of artificial neural networks (ANNs) in pattern recognition and dimensional reduction to construct an efficient and versatile ansatz for the quantum state.  
Importantly, the universal approximation theorems for ANNs guarantee that any wave function can be expressed, provided the NQS ansatz is sufficiently large \cite{Cybenko1989,Hornik1991}. Therefore, the accuracy of NQS simulations can be systematically controlled by varying the ansatz size \cite{chen2024empowering}.  

In this work we represent the impurity wave function as \( \psi_{\pmb{\theta}}(\mathbf{s}) = \exp\big[\chi_{\pmb{\theta}}(\mathbf{s}) + i\,\phi_{\pmb{\theta}}(\mathbf{s})\big]\), where \(\chi_{\pmb{\theta}}(\mathbf{s})\) encodes the (log-)amplitude and \(\phi_{\pmb{\theta}}(\mathbf{s})\) the phase. Despite the universal approximation property, the choice of architecture for \(\chi_{\pmb{\theta}}\) and \(\phi_{\pmb{\theta}}\) plays a crucial role in practice. We realize \(\chi_{\pmb{\theta}}\) with the recently introduced \rwkv{} architecture \cite{peng2023rwkv}, which combines the sampling efficiency of recurrent neural networks with the long-range modeling capabilities of Transformers while avoiding the costly attention mechanism \cite{vaswani2017attention,hibat2020recurrent,sharir2020deepautoregressive}. This makes the \rwkv{} well suited for the irregular connectivity of multi-orbital impurities.
In particular, \rwkv{} defines \(\chi_{\pmb{\theta}}\) in an \emph{autoregressive} manner, so that the corresponding NQS can be sampled sequentially from local conditional probabilities \cite{hibat2020recurrent,sharir2020deepautoregressive}.
This is crucial for SCOL, because it allows us to draw independent identically distributed configurations not only from \(|\psi_{\pmb{\theta}}(\mathbf{s})|^2\) but also from the modified distributions \(p_{a}(\vek{s})\) without computational overhead, at cost \(\mathcal{O}(L)\), where \(L\) is the total system.

We further enhance its performance through \textit{site patching} (grouping adjacent sites into composite units) \cite{spragueVariationalMonteCarlo2024} and by enforcing global charge and magnetization conservation \cite{malyshev2023autoregressive} (see \capp{app:qn-ar} for more details).

\begin{figure}[t]
    \centering
	\includegraphics[width=1.\columnwidth{},trim=0cm .2cm 0.5cm 1.5cm,clip]{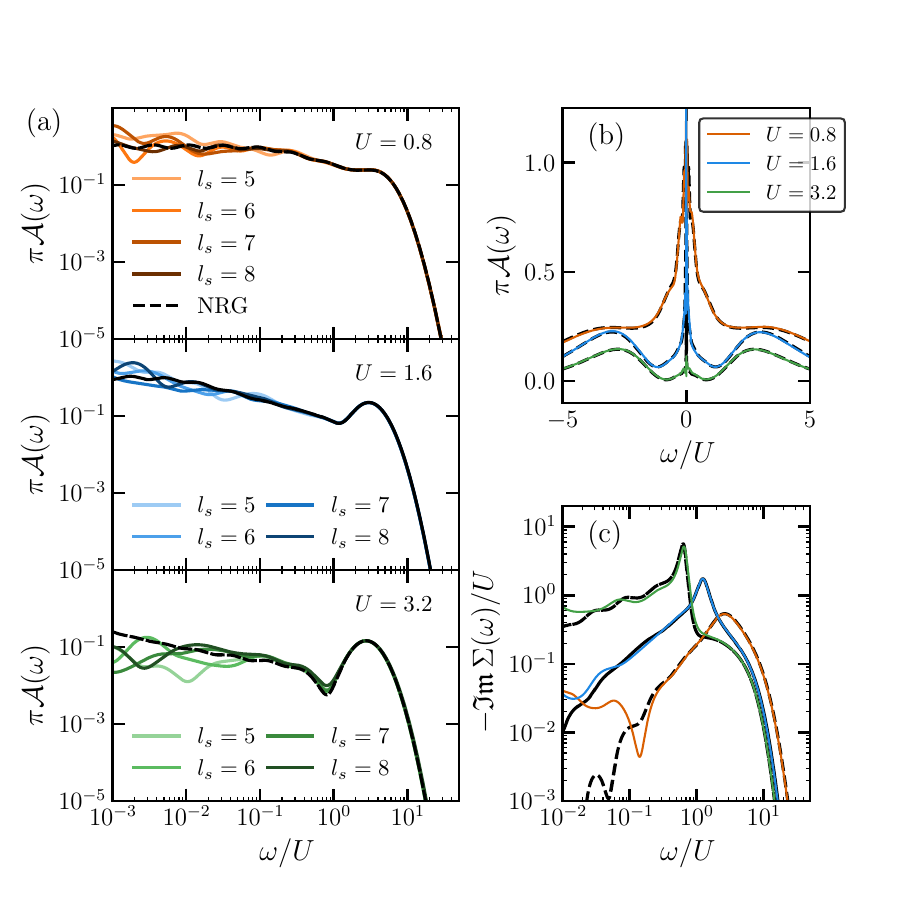}
    \vspace{-0.6cm}
	\caption{Single-impurity Anderson model benchmark for the NQS+SCOL impurity solver. 
 Panel (a) shows the log--log spectral function \(\pi \mathcal{A}(\omega)\) for three interaction strengths \(U=0.8, 1.6, 3.2\) and for several commutator segment depths \(l_s\), compared to NRG reference data (black line). The SCOL set for all \(l_s\) has \(N_s = L_c/l_s\) and \(N_{ab}=2^{14}\) samples are used in \ceqn{eq:ONestimation}. The corresponding spectra on a linear frequency axis \(\omega/U\) are shown in panel (b), highlighting the quasiparticle peak and the Hubbard side bands. In this panel we use \(l_s=8\) for \(U=0.8\) and \(U=1.6\), and \(l_s=7\) for \(U=3.2\). The imaginary part of the self-energy is presented in panel (c), \(-\mathfrak{Im}\,\Sigma(\omega)/U\) again compared to NRG. The bath is a semi-circular band of half-bandwidth \(D=2\), a Wilson discretization parameter \(\Lambda=3\) and a Wilson chain of length \(L_c=35\).}
 \vspace{-.5cm}
	\label{fig:siam}
\end{figure}

\begin{figure}[t]
	\centering
	\includegraphics[width=1\columnwidth]{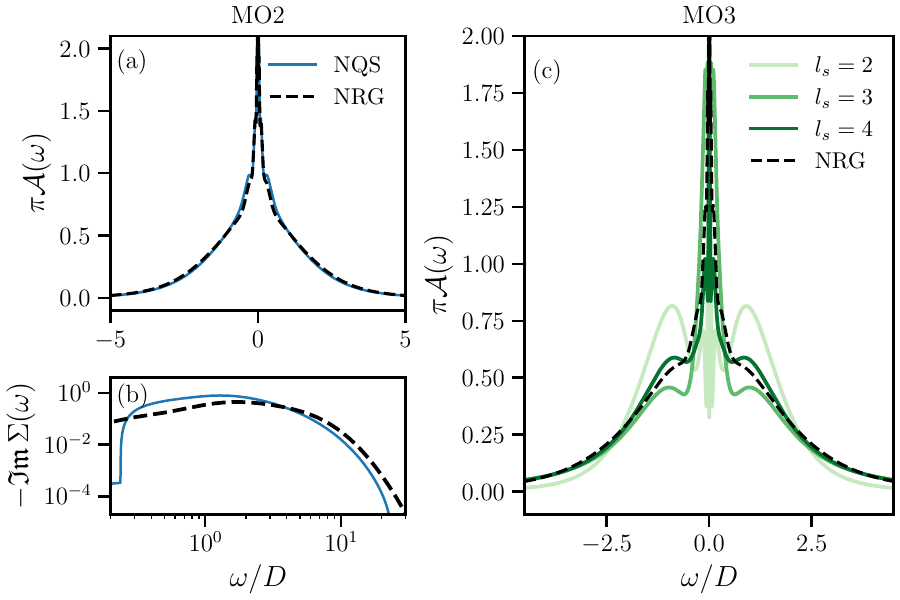}
		\caption{ Panel (a) shows the spectral function \(\pi \mathcal{A}(\omega)\) for the two-orbital (MO2) Hubbard--Kanamori model, obtained with NQS+SCOL and compared to NRG. The imaginary part of the MO2 self-energy, \(-\mathfrak{Im}\,\Sigma(\omega)\), is shown in panel (b). All calculations for MO2 were done for \(\Lambda=3\) with chain length \(L_c=29\), segment depth \(l_s=5\), \(N_s = L_c/l_s\) and \(N_{ab} = 2^{12}\). Panel (c) shows the spectral function \(\pi \mathcal{A}(\omega)\) for the three-orbital Hubbard--Kanamori impurity (MO3) for segment depths \(l_s = 2,3,4\), \(N_s = L_c/(l_s+2)\) and \(N_{ab} = 2^{12}\), compared to the NRG reference with \(L_c = 31\).}
	\label{fig:mo}
    \vspace{-.5cm}
\end{figure}

Capturing the fermionic sign structure of the wave function is a known bottleneck, commonly addressed by augmenting the ansatz with determinants to enforce antisymmetry \cite{luo2019backflow,robledo2022fermionic,rothSuperconductivityTwodimensionalHubbard2025}. To avoid the associated \(\mathcal{O}(L^3)\) evaluation cost, we introduce the \textit{combinatorial phase map} (CPM) layer, detailed in \capp{sec:nqs_details}, which circumvents the high cost of determinants. Nevertheless, the CPM layer retains the characteristic ``small change in, big change out'' sensitivity of determinant-based phases by selecting phase contributions conditioned on the local occupations \(\mathbf{s}\) and summing them up.

\textit{Results}---%
We now benchmark the NQS+SCOL impurity solver first on the single-impurity Anderson model (SIAM), then on the two- (MO2) and three-orbital (MO3) Hubbard--Kanamori impurity models \cite{kugler2019orbital} and compare to \textit{numerical renormalization group} (NRG)  results \cite{bulla1999zero,wilson1975renormalization}. All models use a semi-circular conduction band of half-bandwidth \(D\) for all baths, which is discretized using a discretization parameter \(\Lambda > 1\), following Ref. \cite{bulla2008numerical} to obtain a \textit{Wilson chain} of length \(L_c\).

\emph{Single-impurity Anderson model.} The SIAM Hamiltonian is defined as 
\begin{equation}
\begin{aligned}
\label{eq:aim}
\hat{H}_{\text{SIAM}} &= 
U_d \hat{n}_{d\uparrow}\hat{n}_{d\downarrow}
+ \epsilon_d \sum_\sigma \hat{d}_\sigma^\dagger \hat{d}_\sigma \\
&\quad + 
V\sum_\sigma \left(\hat{d}_\sigma^\dagger \hat{c}_{0\sigma}
+ \hat{c}^\dagger_{0\sigma} \hat{d}_\sigma \right)
+ \hat{H}_{\text{bath}} ,
\end{aligned}
\end{equation}
where we work at half filling with \(U_d/D = 1.47\,U\) and \(\epsilon_d=-U_d/2\). For this benchmark we evaluate \(U = 0.8,~1.6,~3.2\), for which our NQS found ground states with a relative energy error \(\delta E_0 \sim 10^{-6}\) compared to NRG at \(\Lambda=3\) and chain length \(L_c=35\).

Panels (a) and (b) of \cfig{fig:siam} display the spectral function \(\pi \mathcal{A}(\omega)\) for the three \(U\) values and multiple segment depths \(l_s\). The values \(l_s=5,6,7,8\) correspond to \(N_{\rm poles} = 578,~968,~1414,~2274\) poles in the un-broadened \(\mathcal{A}(\omega)\). Even for small \(l_s\), the quasiparticle peak and Hubbard bands appear correctly. Increasing \(l_s\) progressively suppresses oscillations and brings the mid- and high-frequency tails, as well as the Kondo resonance height, into quantitative agreement with NRG.

Panel~(c) shows the self-energy, obtained as
\(\Sigma(\omega) = \tfrac{F(\omega)}{G(\omega)}\),
where \(G(\omega) = \langle\!\langle \hat{c}_j ; \hat{c}_j^\dagger \rangle\!\rangle_{\omega^+}\) is the usual impurity GF and \(F(\omega) = \langle\!\langle \hat{c}_j ; f^\dagger \rangle\!\rangle_{\omega^+}\) with
\(\hat{f}^\dagger = [\hat{H}_{\rm int},\hat{c}_j^\dagger]\) and \(\hat{H}_{\rm int}\) the quartic part of \(\hat{H} \) \cite{bulla2008numerical}. By construction, this commutator operator \(\hat{f}^\dagger\) is always contained in the SCOL operator set, so both correlators \(F(\omega)\) and \(G(\omega)\) are obtained from the \emph{same} SCOL operator space and thus from the same generalized eigenproblem. As a result, evaluating \(\Sigma(\omega)\) is essentially cost-free. The SCOL self-energy closely matches NRG over several decades before breaking down at the lowest frequencies.

The location of the low-frequency breakdown is governed by two contributions. In the limit of infinite samples there is no Monte Carlo error in the matrices \(\mat{G}\) and \(\mat{H}\) such that the smallest resolvable frequencies are set by the deviation of the NQS ground state \(\ket{\psi_{\boldsymbol{\theta}}}\) from the exact one. For finite sample sizes (finite Monte Carlo error), noise in the matrix elements of \(\mat{H}\) and \(\mat{G}\) induces fluctuations in the smallest eigenvalues and therefore an effective infrared cutoff \(\omega_{\mathrm{noise}} > \delta E_0\). Larger \(l_s\) increases the expressiveness of the operator set and narrows intrinsic linewidths, but also increases sensitivity to sampling noise. We find that for a fixed sample budget there is a broad window of \(l_s\) where the spectra are smooth and extremely close to NRG. Outside this window small \(l_s\) leads to truncation errors and large \(l_s\) amplifies sampling noise. This can be seen in panel (a) for \(U = 3.2\), as \(l_s = 7\) adheres better to the NRG reference than \(l_s = 8\).

For MO2 and MO3 we benchmark the multi-orbital Hubbard--Kanamori model \cite{kugler2019orbital}
\begin{align}
\label{eq:hubkan}
&\hat{H} = 
\sum_{m} \hat{H}_{\text{SIAM},m}
+ (U - 2 J) \sum_{m \neq m'} \hat{n}_{m\uparrow} \hat{n}_{m' \downarrow} \\
&\quad + (U - 3 J) \sum_{m < m',\sigma} \hat{n}_{m\sigma} \hat{n}_{m'\sigma}  \nonumber\\
&\quad - J  \sum_{m \neq m'} 
\hat{d}^\dagger_{m \uparrow} \hat{d}_{m \downarrow}
\hat{d}^\dagger_{m' \downarrow} \hat{d}_{m' \uparrow} \nonumber
+ J  \sum_{m \neq m'}
\hat{d}^\dagger_{m \uparrow} \hat{d}^\dagger_{m \downarrow}
\hat{d}_{m' \downarrow} \hat{d}_{m' \uparrow} ,\nonumber
\end{align}
with parameters \(U = 2/D\), \(J = 0.3/D\), \(\epsilon_d = -\big[U/2 + (n_b-1)(U-\tfrac{5}{2}J)\big]\) and \(n_b\) the number of baths (or orbitals). All calculations use a Wilson discretization \(\Lambda=3\) and chain lengths \(L_c=29\) for \(n_b = 2\) and \(L_c=31\) for \(n_b = 3\). For these parameters, the NQS ground-state optimization yields an energy variance of \(\mathbb{V}[E_0] = 2\times 10^{-5}\) for MO2 and \(\mathbb{V}[E_0] = 10^{-3}\) for MO3.

For the calculation of the MO2 spectral function we use \(l_s = 5\) with \(N_s = L_c / l_s\), and for MO3 we use \(l_s = 2,3,4\) with \(N_s = L_c / (l_s+2)\), which in both cases yields \(N_{\rm poles} \sim 2600\) for the highest \(l_s\).

For the two-orbital impurity (MO2), \cfig{fig:mo} (a) demonstrates that the NQS solver quantitatively reproduces the quasiparticle peak, the onset and position of the Hubbard bands, and the high-frequency tails of \(\mathcal{A}(\omega)\). Small deviations at very low frequencies correlate with the point where \(-\mathfrak{Im}\,\Sigma(\omega)\) in panel~(b) starts to strongly drift away from the NRG result, signaling the same sampling-limited infrared cutoff as in the SIAM.

For the three-orbital impurity (MO3), \cfig{fig:mo} (c) compares NQS spectra for segment depths \(l_s = 2,3,4\) to the NRG reference. 
At \(l_s=2\) strongly deviates from the reference, but increasing \(l_s\) systematically improves the line shape and brings the Hubbard bands closer to the NRG result. A modest overestimation of the Hubbard-band peak height and a suppression of spectral weight below \(|\omega|/D \sim 10^{-1}\) remain, consistent with the finite number of poles and the ground-state accuracy reached in this calculation. The clear trend with \(l_s\) indicates that a further, moderate increase (e.g. \(l_s=5\)) would likely remove most of the remaining discrepancy, while the present data already demonstrate that the SCOL construction remains robust and scalable in the three-orbital case.

\textit{Discussion}---%
We have demonstrated that a combination of an autoregressive NQS ground state, a low-cost fermionic phase representation via the combinatorial phase map, and the segmented commutator operator-Lanczos construction yields a flexible real-frequency impurity solver. Once the SCOL matrices are available, both the GF \(G(\omega)\) and the correlator \(F(\omega)\) entering the self-energy \(\Sigma(\omega)=F(\omega)/G(\omega)\) are obtained from a single generalized eigenvalue problem, which makes access to self-energies essentially free of cost. The quality of the spectra is controlled by two independent knobs: the SCOL parameters \((N_s,l_s,m_{\max})\), which govern truncation errors, and the number of Monte Carlo samples, which controls the low-frequency noise floor. In addition, both the NQS ground-state optimization and the SCOL estimation of the projected matrices exhibit a high degree of parallelism, making the full workflow naturally suited to modern GPU and multi-node architectures. From a practical perspective, the method is particularly attractive as an ingredient for DMFT and related embedding schemes. The impurity graphs generated in realistic multi-orbital DMFT, including cluster DMFT setups with several correlated sites or orbitals, can be treated without redesigning the solver, and the same NQS and SCOL machinery applies equally to single-site, multi-orbital, and cluster geometries \cite{maier2005quantum}. The benchmarks presented here suggest that, with moderate segment depths and sample numbers, one can already achieve real-frequency GFs and self-energies of a quality comparable to NRG for up to three orbitals. We therefore expect that NQS-based real-frequency impurity solvers will become a valuable complement to existing impurity solvers, especially in regimes where sign problems or entanglement growth limit other methods, and where scalable, hardware-efficient parallelization is crucial.

\begin{acknowledgments}
The authors acknowledge insightful discussions with Wladislaw Krinitsin, Seung-Sup Lee, Mathias Steinhuber, Oleksii Malyshev, Antoine Georges, and Sudeshna Sen. 
The implementation of the NQS ground-state optimization and the SCOL procedure was facilitated by the \emph{jVMC} library \cite{schmittJVMCVersatilePerformant2022}.
This work was supported via the Helmholtz Initiative and Networking Fund, grant no.~VH-NG-1711.
The authors gratefully acknowledge the Gauss Centre for Supercomputing e.V. (www.gauss-centre.eu) for funding this project by providing computing time through the John von Neumann Institute for Computing (NIC) on the GCS Supercomputer JUWELS at Jülich Supercomputing Centre (JSC)
and the scientific support and HPC resources provided by the Erlangen National High Performance Computing Center (NHR@FAU) of the Friedrich-Alexander-Universität Erlangen-Nürnberg (FAU) under the NHR project b245da / JA-22555. NHR funding is provided by federal and Bavarian state authorities. NHR@FAU hardware is partially funded by the German Research Foundation (DFG) – 440719683. \\
\textit{Data Availability}---%
The data and code used to produce the presented plots are available at Ref.~\cite{rigo_2025_17865207}.
\end{acknowledgments}

\begin{appendix}

\section{Spectrum invariance under linear transformations}
\label{sec:invar_proof}

We show the invariance of the spectrum under linear transformation of the operator set by showing the respective transfromations of the generalized eigenvalue problem.
Let \(\{\hat O_i\}_{i=1}^N\) be a set of operators and define the respective span
\begin{equation}
  \ket{\phi_i} = \hat O_i \ket{\psi_0}, \qquad
  \mathcal{S} = \mathrm{span}\{\ket{\phi_i}\}_{i=1}^N \;,
\end{equation}
with the seed state \(\ket{\psi_0}\). For coefficients \(\boldsymbol{\theta} = (\theta_1,\dots,\theta_N)^{T}\) a state in the span is given as
\begin{equation}
  \ket{\psi(\boldsymbol{\theta})} = \sum_{i=1}^N \theta_i \ket{\phi_i} \in \mathcal{S} \;.
\end{equation}
To get an eigenstates of the effective Hamiltonian projected in to the span one solves the generalized eigenvalue problem
\begin{equation}
  \mat{H} \boldsymbol{\theta} = E\, \mat{G} \boldsymbol{\theta},
\end{equation}
with matrices
\begin{equation}
  \mat{H}_{ij} = \bra{\phi_i}\hat H\ket{\phi_j},
  \qquad
  \mat{G}_{ij} = \braket{\phi_i|\phi_j}.
\end{equation}

Now take any invertible matrix \(\mat{M} \in \mathbb{C}^{N\times N}\) and define a new
operator set
\begin{equation}
  \hat{\tilde O}_a = \sum_{i=1}^N \mat{M}_{ai} \hat O_i,
  \qquad
  \ket{\tilde\phi_a} = \hat{\tilde O}_a \ket{\psi_0}
                     = \sum_{i=1}^N \mat{M}_{ai} \ket{\phi_i}.
\end{equation}
The corresponding matrices are
\begin{align*}
  \tilde{\mat{H}}_{ab}
   &= \bra{\tilde\phi_a}\hat H\ket{\tilde\phi_b}
    = \sum_{ij} \mat{M}_{ai} \mat{H}_{ij} \mat{M}^{*}_{bj}
    = (\mat{M} \mat{H} \mat{M}^\dagger)_{ab}, \\
  \tilde{\mat{G}}_{ab}
   &= \braket{\tilde\phi_a|\tilde\phi_b}
    = \sum_{ij} \mat{M}_{ai} \mat{G}_{ij} \mat{M}^{*}_{bj}
    = (\mat{M} \mat{G} \mat{M}^\dagger)_{ab}.
\end{align*}
If the solution fo the eigenvalue problem is \(\mat{H}\boldsymbol{\theta}=E \mat{G}\boldsymbol{\theta}\), then set
\(\tilde{\boldsymbol{\theta}} = (\mat{M}^{-1})^{\dagger} \boldsymbol{\theta}\). With this ansatz we find
\begin{align}
  \tilde{\mat{H}} \tilde{\boldsymbol{\theta}}
  &= \mat{M} \mat{H} \mat{M}^\dagger (\mat{M}^{-1})^{\dagger}  \boldsymbol{\theta}
  = \mat{M} \mat{H} \boldsymbol{\theta} \\
  & = E \mat{M} \mat{G} \boldsymbol{\theta}
  = E \mat{M} \mat{G} \mat{M}^\dagger (\mat{M}^{-1})^{\dagger}  \boldsymbol{\theta}
  = E \tilde{\mat{G}} \tilde{\boldsymbol{\theta}}.
\end{align}
Thus \(E\) is also an eigenvalue of the transformed problem
\(\tilde{\mat{H}} \tilde{\boldsymbol{\theta}} = E \tilde{\mat{G}} \tilde{\boldsymbol{\theta}}\) and the state is unchanged:
\begin{equation}
  \ket{\psi(\boldsymbol{\theta})}
  = \sum_i \theta_i \ket{\phi_i}
  = \sum_a \tilde\theta_a \ket{\tilde\phi_a}.
\end{equation}
Hence, both the effective energies and the effective states are invariant under 
linear transformation of the operator set \(\{\hat O_i\}\).

\section{Details of NQS architecture and ground-state search}
\label{sec:nqs_details}

\subsection{NQS architecture}
\label{app:nqs_architecture}

\begin{figure}[t]
	\centering
	\includegraphics[width=1\columnwidth]{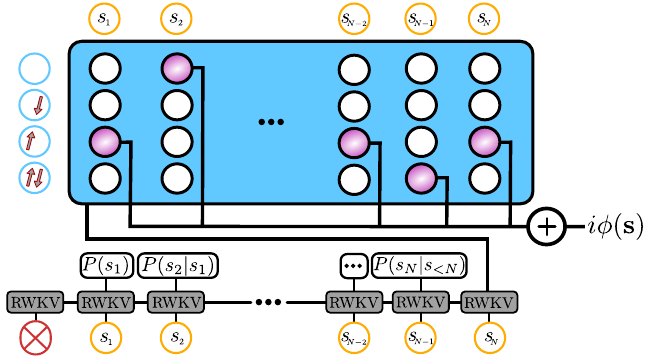}  
	\caption{Schematic representation of the combinatorial phase map (CPM) that supplies the complex phase of the NQS wave function.}
	\label{fig:phase_layer}
	\vspace{-0.5cm}
\end{figure}

As outlined in the main text, the core architecture of the NQS for both the SIAM and the multi-orbital (MO2, MO3) models is provided by the \rwkv{} network \cite{peng2023rwkv}.  
The \rwkv{} is fully specified by four hyperparameters:
\begin{itemize}
\item The number of layers or \textit{depth} (\(\mathrm{Depth}\)), which determines how many \textit{time-mixing} and \textit{channel-mixing} blocks are concatenated.

\item The \textit{embedding dimension} (\(d_e\)), which maps each input \(s_i \in\{\ \downarrow,\ \varnothing,\ \uparrow,\ \uparrow\downarrow\ \}^\pi\) to an initial vector
\(\mathbf{x}^{(0)}\in \mathbb{R}^{d_e}\) fed into the first \rwkv{} block, where \(\pi\) is the \textit{patch size}.  
The patch size specifies how many local Hilbert spaces are combined into a single composite site, so that the effective system size is \(P=L/\pi\).

\item The \textit{hidden layer dimension} (\(d_h\)), which sets the size of the residual blocks and thus the width of the internal representations in each layer.

\item The number of \textit{attention heads} (\(n_H\)), which is the number of parallel \rwkv{} subunits that each learn complementary features of the input sequence.
\end{itemize}

We parameterize the complex NQS as \(\psi_{\pmb{\theta}}(\mathbf{s}) = \exp\big[\chi_{\pmb{\theta}}(\mathbf{s}) + i\,\phi_{\pmb{\theta}}(\mathbf{s})\big]\), where the real-valued log–amplitude \(\chi_{\pmb{\theta}}(\mathbf{s})\) is provided by the RWKV and 
the task of \(\phi_{\pmb{\theta}}(\mathbf{s})\) is encoding the fermionic sign structure, which is a well-known bottleneck.
A common solution is to augment the ansatz with determinants or \textit{Pfaffians} to enforce antisymmetry and to smoothen the energy landscape \cite{luo2019backflow,robledo2022fermionic,rothSuperconductivityTwodimensionalHubbard2025}, at the price of a \(\mathcal{O}(L^3)\) network evaluation cost.

Here, we introduce an alternative building block for fermionic wave functions -- the \textit{combinatorial phase map} (CPM), which circumvents the high cost of determinants. On top of the real-valued \rwkv{} backbone, with final hidden state \(\mathbf{x}^{(L)} \in \mathbb{R}^{d_h}\), we add a single complex-valued phase layer that defines
\begin{equation}
\label{eq:cpm_definition}
  \phi_{\pmb{\theta}}(\mathbf{s})
  = \sum_i \left( b_{ij} + \sum_k W_{ijk} x^{(L)}_k \right)\big\vert_{j = s_i} ,
\end{equation}
where \(\mat{W} \in \mathbb{R}^{4 L \times d_h}\) and \(\mathbf{b} \in \mathbb{R}^{4 L}\) are the variational weights and biases of the CPM layer, and \(s_i \in \{1,2,3,4\}\) encodes the local occupation at site \(i\). In \ceqn{eq:cpm_definition} the phase is built by selecting, for each site, one of four site- and configuration-dependent phase channels and summing their contributions.
Despite its reduced cost, the CPM layer achieves a similar “small change in, big change out” effect as determinant-based phases, as illustrated in \cfig{fig:phase_layer}, by making the global phase highly sensitive to local changes in \(\mathbf{s}\).
To reliably induce such sharp phase variations, the bias tensor \(\mathbf{b}\) is initialized with a large variance. 

While the CPM phase layer saves on network evaluation time, it has a memory cost that scales linearly with system size, \(\propto \mathcal{O}(L)\). This cost can be capped by restricting the CPM dependence to a fixed number of sites \(N_{\rm cap}\), which we choose to be the impurity sites and an equal number of bath sites attached to each orbital. In this case, the CPM parameters become \(\mat{W} \in \mathbb{R}^{4 N_{\rm cap} \times d_h}\) and \(\mathbf{b} \in \mathbb{R}^{4 N_{\rm cap}}\), so that the memory cost scales as \(\mathcal{O}(N_{\rm cap})\) instead of \(\mathcal{O}(L)\). In our tests, this controlled reduction of expressivity does not noticeably degrade performance.

It is important to note that not all phases need to be learned variationally. For the semi-infinite baths, phase changes associated with fermionic exchange can be handled trivially by mapping the occupation configuration \(\mathbf{s}\) to a bit string and assigning a sign based on the parity of fermion permutations. In the single-orbital SIAM this simple construction is sufficient to render the problem stoquastic.  
In addition, we obtain improved ground-state energies and observables by enforcing symmetries directly in the ansatz. All models are at half filling and thus, we impose particle--hole and spin-flip symmetry on the NQS \cite{chooSymmetriesManyBodyExcitations2018}.
For multi-orbital systems with degenerate orbitals, we further enforce simple Abelian orbital symmetries, that are permutations of equivalent orbitals. Beyond enforcing the symmetries on the level of wave function amplitudes we use quantum-number--conserving autoregressive sampling as introduced in Ref.~\cite{malyshev2023autoregressive}. Our specific implementation is outlined in \capp{app:qn-ar}.

In Tab.~\ref{tab:nqs_p} we summarize the key \rwkv{} and CPM hyperparameters used for the SIAM and Hubbard--Kanamori NQS models.

\begin{table}
	\centering
	\begin{tabular}{c||c|c|c}
     	& SIAM & MO2 & MO3 \\
     	\hline\hline
    	\(\mathrm{Depth}\) & 4 & 4 & 4\\
    	\(d_e\) & 32 & 80 & 70\\
    	\(d_h\) & 64 & 64 & 32 \\
    	\(n_H\) & 1 & 1 & 1 \\
        \(\pi\) & 2 & 4 & 2\\
    	channel mixing & No & No & No \\
    	\(N_{\rm cap}\) for CPM & \o & 6 & 6 \\
        \(\mat{Z}_2\) spin & Yes & Yes & Yes \\
        \(\mat{Z}_2\) charge & Yes & Yes & Yes \\
        Orbital symmetry & \o & Yes & No \\
	\end{tabular}
	\caption{RWKV+CPM parameters.}
	\label{tab:nqs_p}
\end{table}

\subsection{NQS ground-state search}
\label{app:nqs_gss}

The quintessential idea for how to attain the ground state of a Hamiltonian \(\hat{H}\) with an NQS \(\ket{\psi_{\pmb{\theta}}}\) is captured by the variational principle
\begin{equation}
\label{eq:loss}
\pmb{\theta}^* = \arg \min_{\pmb{\theta}}
\bra{\psi_{\pmb{\theta}}}\hat{H}\ket{\psi_{\pmb{\theta}}} \;,
\end{equation}
which here was rearranged to solve for the parameters that let \(\ket{\psi_{\pmb{\theta}}}\) closest approximate the ground state. Note that no normalization is needed for the Rayleigh quotient, since \(\psi_{\pmb{\theta}}\) is autoregressive.

The optimization problem implied by Eq.~\eqref{eq:loss} is non-convex, so the global minimum is not known in practice. A particularly effective optimizer in this setting is \textit{stochastic reconfiguration} (SR), a second-order gradient-based method that approximates imaginary-time evolution in parameter space. For all ground-state searches we use a variation of SR, called \textit{minSR} \cite{chen2024empowering}. Unlike SR, minSR limits the number of samples used to estimate the variational energy \(E_{\pmb{\theta}} = \bra{\psi_{\pmb{\theta}}}\hat{H}\ket{\psi_{\pmb{\theta}}}\) and its gradient, rather than the number of variational parameters. We use the particular realization of minSR presented in Ref.~\cite{rendeSimpleLinearAlgebra2024b}, with the regularization procedure introduced in Ref.~\cite{goldshlagerKaczmarzinspiredApproachAccelerate2024}. For the ground-state searches presented here we choose a regularization strength \(\lambda = 10^{-4}\), perform \(n_{\rm GD} = 2\times 10^{4}\) SR steps with learning rate \(\eta = 0.05\) and apply adaptive gradient clipping such that the gradient norm never exceeds the running average by more than a factor of \(10\). To estimate \(E_{\pmb{\theta}}\) and \(\nabla_{\pmb{\theta}}E_{\pmb{\theta}}\) we use \(2^{12}\) samples, which are drawn autoregressively, so no additional parameters are required to specify the sampling process.

\section{Details of the calculation of spectral functions}
\label{sec:gfcalc}

To compute zero-temperature GFs such as
\(\mathcal{G}_{ij}(\omega) = \zuba{ \hat{c}_{i};\hat{c}^\dagger_{j}}_{\omega^+}\),
one can consider many-body eigenstates
\(\{\ket{\psi^n_{Q,m_z}}\}\) of \(\hat{H}\), labeled by abelian charge \(Q\) and magnetization \(m_z\) quantum numbers.
Denoting by \(Q^0\) and \(m_z^0\) the charge and magnetization quantum numbers of the ground state
\(\ket{\psi^0_{Q^0,m^0_z}}\), the zero-temperature GF can be written in the
\textit{Lehmann representation} as
\begin{widetext}
\begin{equation}
\label{eq:lehmann}
\mathcal{G}_{ij}(\omega)
= 
\sum_n
\frac{
  \bra{\psi^0_{Q^0,m^0_z}}\hat c^\dagger_{i\sigma}\ket{\psi^n_{Q^0 - 1,m^0_z - \sigma}}
  \bra{\psi^n_{Q^0 - 1,m^0_z - \sigma}}\hat c_{j\sigma}\ket{\psi^0_{Q^0,m^0_z}}
}{
  \omega^+ - E^0 + E^n_{Q^0 - 1,m^0_z - \sigma}
}
+
\sum_n
\frac{
  \bra{\psi^0_{Q^0,m^0_z}}\hat c_{i\sigma}\ket{\psi^n_{Q^0 + 1,m^0_z + \sigma}}
  \bra{\psi^n_{Q^0 + 1,m^0_z + \sigma}}\hat c^\dagger_{j\sigma}\ket{\psi^0_{Q^0,m^0_z}}
}{
  \omega^+ + E^0 + E^n_{Q^0 + 1,m^0_z + \sigma}
}\;. 
\end{equation}
\end{widetext}
In contrast, SCOL does not rely on a set of orthogonal, normalized eigenstates, but on the
non-orthogonal states \(\ket{\phi_a} = \hat{O}_a\ket{\psi^0_{Q^0,m^0_z}}\) that span an operator-generated subspace.

Given the SCOL operator set \(\mathcal{B}\), we construct the projected energy matrix \(\tilde{\mathbb{H}}\) and overlap matrix \(\mat{G}\) for particle and hole excitations according to Eq.~\eqref{eq:scol_mats}, using the seed operators \(\hat{O}^p_0=\hat{c}_j^\dagger\) and \(\hat{O}^h_0=\hat{c}_j\), respectively.  
To obtain the spectral poles, we use the generalized Lehmann representation \cite{soriano2014theory,rosenberg2022fermi,*rosenberg2023dynamical}, which follows from the generalized eigenvalue problem in Eq.~\eqref{eq:gen_eigp}.  
For particle (\(+\)) and hole (\(-\)) contributions, the spectral function can be written in compact matrix form as
\begin{align}
\label{eq:luss_lehm}
&\mathcal{A}_{ij}(\omega) = -\tfrac{1}{\pi}\imag \mat{G}\cdot\big[(\omega^+ \pm E_0)\cdot\mat{G} \mp \mat{H}\big]^{-1}\cdot\mat{G} \\
&= \sum_\nu \big[\sqrt{\mat{G}}\cdot\mathbb{U}\big]_{i\nu} \cdot \big[\mathbb{U}^\dagger \cdot\sqrt{\mat{G}}\big]^T_{\nu j}\times\delta(\omega \pm E_0\mp D_\nu) \;. \nonumber
\end{align}
where the second line uses the Cauchy principal-value formula to obtain the spectral function as a sum over poles.

In deriving Eq.~\eqref{eq:luss_lehm}, we use the identity
\begin{equation}
\mat{G}^{-1/2}\,{\mathbb{H}}\,\mat{G}^{-1/2}
= \mathbb{U}^\dagger
  \operatorname{diag}(D_1,D_2,\dots,D_{N_{\rm poles}})
  \mathbb{U} ,
\end{equation}
with \(\mathbb{U}\) the eigenbasis of \({\mathbb{H}}\) and \(\{D_\nu\}\) the generalized eigenvalues, which we denote with \(D_\nu\) since they are obtained in a different quantum number subspace than the ground state.  
According to the Cauchy principal-value construction, the pole positions \(\omega_\nu = \pm E_0\mp D_\nu\) are therefore given by the generalized eigenvalues, while the corresponding spectral weights are encoded in the eigenvectors via the products
\begin{equation}
w_\nu^{(ij)} =
\big[\sqrt{\mat{G}}\,\mathbb{U}\big]_{i\nu}
\big[\mathbb{U}^\dagger\sqrt{\mat{G}}\big]_{\nu j}.
\end{equation}
This fully specifies the discrete spectral functions \(\mathcal{A}_{ij}(\omega)\) through
\(\mathcal{A}_{ij}(\omega) = -\frac{1}{\pi}\,\mathfrak{Im}\,\mathcal{G}_{ij}(\omega)\).  
The resulting discrete spectra are finally rendered continuous by applying the broadening procedure of Ref.~\cite{weichselbaum2007sum}, originally developed for numerical renormalization group calculations.

\section{Quantum-number--conserving autoregressive sampling}
\label{app:qn-ar}

In the following, we outline our procedure for exclusively drawing samples with fixed target quantum numbers for charge (\(Q^0\)) and magnetization (\(M^0\)). The method follows the prescription of Ref.~\cite{malyshev2023autoregressive} and incorporates operator-sequence constraints at every step of the sampling process, which allow us to apply operators from the SCOL set to the state directly at sampling time, while also patching sites.

\subsubsection{Local Hilbert space and quantum numbers}

The multi-orbital quantum impurity model graph can always be organized into a linear chain, despite some links get non-local connectivity, this is a worthwhile trade-off for the autoregressive process to be feasible. Therefore we consider a one-dimensional lattice of \(L\) sites with spinful fermions. At each site \(i\in\{1,\dots,L\}\) we use the local Hilbert alphabet
\begin{equation}
  \mathcal{A}_1
  = \{\ \downarrow,\ \varnothing,\ \uparrow,\ \uparrow\downarrow\ \}
  = \{0,1,2,3\},
\end{equation}
where the last equality indicates a fixed integer encoding.

Local charge or electron (\(Q\)) and magnetization (\(M\)) contributions from each local Hilbert space configuration are 
\begin{align}
  Q(\downarrow)=1,~ & Q(\varnothing)=0,~ Q(\uparrow)=1,~ Q(\uparrow\downarrow)=2, \\
  M(\downarrow)=-1,~ & M(\varnothing)=0,~ M(\uparrow)=1,~ M(\uparrow\downarrow)=0.
\end{align}
For a configuration from the occupation basis \(\vek{x}=(x_1,\dots,x_L)\in\mathcal{A}_1^{\times L}\) the total charge and magnetization is
\begin{equation}
  Q(\vek{x}) = \sum_{i=1}^L Q(x_i), \qquad
  M(\vek{x}) = \sum_{i=1}^L M(x_i).
\end{equation}
The target symmetry sector is
\begin{equation}
  \mathcal{X}_{Q^0,M^0}
  = \{\, \vek{x}\in\mathcal{A}^{\times L}_1 : Q(\vek{x})=Q^0,\ M(\vek{x})=M^0 \,\}.
\end{equation}

\subsubsection{Operator-sequence constraints}

To efficiently estimate the \(\mat{G}\) matrix we sample directly \(p_a(\vek{x}) \propto |\bra{\vek{x}}\hat{O}_a\ket{\psi_{\pmb{\theta}}}|^2\), where \(\hat{O}_i\) is a product of fermionic creation and annihilation operators. This type of operator can directly be applied to an autoregressive \(\ket{\psi_{\pmb{\theta}}}\) and sampled, since \(\hat{O}_i\) simply enforces local occupation configurations. These per-site restrictions are encoded in an operator sequence
\begin{equation}
  \operatorname{op}
  = \bigl(\operatorname{op}^{\uparrow},\operatorname{op}^{\downarrow}\bigr)
  \in \{0,1,2\}^{\times L} \times \{0,1,2\}^{\times L},
\end{equation}
where at each site \(i\) and spin flavor \(\sigma\in\{\uparrow,\downarrow\}\):
\begin{itemize}
  \item \(\operatorname{op}^\sigma_i = 0\) enforces the spin-\(\sigma\) orbital to be empty,
  \item \(\operatorname{op}^\sigma_i = 1\) enforces it to be occupied,
  \item \(\operatorname{op}^\sigma_i = 2\) leaves it unconstrained.
\end{itemize}
A local state \(a\in\mathcal{A}_1\) is allowed at site \(i\) if its spin content is compatible with \(\operatorname{op}_i\). We denote the allowed set by \(\mathcal{C}_{\operatorname{op}}(i)\subseteq\mathcal{A}_1\).

\subsubsection{Autoregressive factorization and logit masks}

The RWKV is an autoregressive network, meaning that its respective Born probability distribution \(p_{\pmb{\theta}}(\vek{x}) = |\bra{\vek{x}}\ket{\psi_{\pmb{\theta}}}|^2\) factorizes over configurations as
\begin{equation}
  p_{\pmb{\theta}}(\vek{x})
  = \prod_{t=1}^L p_{\pmb{\theta}}(x_t\mid \vek{x}_{<t}),
  \label{eq:app-ar-factorization}
\end{equation}
where \(\vek{x}_{<t}:=(x_1,\dots,x_{t-1})\) and \(\pmb{\theta}\) are neural-network parameters.
We refer to \(\vek{x}_{<t}\) as the \emph{prefix} of length \(t-1\), that is the partial configuration obtained by fixing all sites strictly to the left of \(t\). The sampling process runs sequentially starting from the empty prefix \(\vek{x}_{<1} = \varnothing\), at step \(t\) the network is evaluated on the current prefix to produce partial probabilities over the local alphabet (which are the configurations of the local Hilbert space of a lattice site), which allows a local state \(x_t\) to be drawn from the resulting conditional distribution
\begin{equation}
  x_t \sim p_{\pmb{\theta}}(x_t\mid \vek{x}_{<t}) \;.
\end{equation}
Repeating this procedure for \(t=1,\dots,L\) yields a full configuration \(\vek{x}\).

In practice the network does not produce probabilities, but at each step \(t\) outputs unnormalized values that we consider logarithms of positive numbers, known as \textit{logits}
\(\ell_{\pmb{\theta}}(t,\cdot\mid \vek{x}_{<t})\in\mathbb{R}^{|\mathcal{A}_1|}\). By applying the \textit{softmax} activation function actual conditioned probabilities can be obtained,
\begin{equation}
  p_{\pmb{\theta}}(\cdot\mid \vek{x}_{<t})
  = \mathrm{softmax}\bigl(\ell_{\pmb{\theta}}(t,\cdot\mid \vek{x}_{<t})\bigr).
\end{equation}

Without constraining \(\ell_{\pmb{\theta}}\) every configuration \(\vek{x}\in\mathcal{A}_1^{\times L}\) can be drawn. To impose hard constraints, we add a \textit{logit mask }\(m_t(\cdot\mid \vek{x}_{<t})\) and instead use
\begin{equation}
  \tilde \ell_{\pmb{\theta}} = \ell_{\pmb{\theta}} + m_t,\qquad
  p_{\pmb{\theta}}(\cdot\mid \vek{x}_{<t}) = \mathrm{softmax}(\tilde \ell_{\pmb{\theta}}).
\end{equation}
The mask decomposes as
\begin{equation}
  m_t(a) = m_t^{\mathrm{op}}(a) + m_t^{\mathrm{QN}}(a),
\end{equation}
where \(m_t^{\mathrm{op}}\) enforces the local operator sequence and
\(m_t^{\mathrm{QN}}\) is a \emph{quantum-number mask} that removes choices incompatible with the global target constraints \((Q^0,M^0)\).

The operator mask is purely local and therefore easy to compute
\begin{equation}
  m_t^{\mathrm{op}}(a) =
  \begin{cases}
    0, & a\in \mathcal{C}_{\operatorname{op}}(t),\\[2pt]
    -\infty, & \text{otherwise}
  \end{cases} \;.
\end{equation}
The quantum-number mask is more complex, since the magnetization is not monotonically increasing, but can also decrease throughout the generation process of a sample \(\vek{x}\).

\subsubsection{Quantum-number subspace restriction}

The quantum-number mask \(m_t^{\mathrm{QN}}\) is built from two tables,
\(M_{\min}(t,q)\) and \(M_{\max}(t,q)\), which encode the achievable magnetization intervals on suffixes for fixed electron counts.

For \(t\in\{0,1,\dots,L\}\) let the \textit{suffix} (tail of the chain after position \(t\)) starting at \(t\) denote the sites \(\{t+1,\dots,L\}\), with \(t=L\) corresponding to the empty suffix. For each integer \(q\in\{0,1,\dots,2L\}\) we define
\begin{align}
  M_{\min}(t,q)
  &= \min \{\, M(y) : y\in\mathcal{A}_1^{L-t},\ Q(y)=q,\nonumber\\[-4pt]
  &\hspace{1.5cm}
     y \text{ allowed by } \operatorname{op} \text{ on } \{t+1,\dots,L\} \,\},
  \label{eq:app-mmin-def}
  \\[2pt]
  M_{\max}(t,q)
  &= \max \{\, M(y) : y\in\mathcal{A}_1^{L-t},\ Q(y)=q,\nonumber\\[-4pt]
  &\hspace{1.5cm}
     y \text{ allowed by } \operatorname{op} \text{ on } \{t+1,\dots,L\} \,\}.
  \label{eq:app-mmax-def}
\end{align}
If no such configuration \(y\) exists, we set \(M_{\min}(t,q)=+\infty\) and \(M_{\max}(t,q)=-\infty\).

For the empty suffix at \(t=L\),
\begin{equation}
  M_{\min}(L,0) = M_{\max}(L,0) = 0,
\end{equation}
and for \(q\neq 0\),
\begin{equation}
  M_{\min}(L,q) = +\infty,\qquad
  M_{\max}(L,q) = -\infty.
\end{equation}
The first condition reflects that zero magnetization can be achieved uniquely by the empty configuration with zero electrons and the second ensures that it is impossible to place electrons on an empty suffix.

\subsubsection{Recursive construction}

The tables \(M_{\min}(t,q)\) and \(M_{\max}(t,q)\) are computed by a backward recursion in \(t\), starting from the boundary conditions at \(t=L\). At a given suffix index \(t\), the next physical site to be filled is \(t+1\). For each allowed local state \(a\in \mathcal{C}_{\operatorname{op}}(t+1)\), any suffix configuration \(y\) on \(\{t+1,\dots,L\}\) formed by placing \(a\) at \(t+1\) and an admissible tail \(y'\) obeys
\begin{equation}
  Q(y) = Q(a) + Q(y'), \qquad
  M(y) = M(a) + M(y').
\end{equation}
Thus, for fixed total electrons \(q\) on the suffix, choosing \(a\) and distributing the remaining \(q' = q - Q(a)\) electrons on the tail yields magnetizations in
\begin{equation*}
   \{M(a) + M_{\min}(t+1,q'),\dots,M(a)+\ M_{\max}(t+1,q')\}.
\end{equation*}
Taking the lower (upper) envelope over all allowed \(a\) gives
\begin{align}
  M_{\min}(t,q)
  &= \min_{a\in \mathcal{C}_{\operatorname{op}}(t+1)}
     \bigl[ M_{\min}(t+1,q-Q(a)) + M(a) \bigr],
  \label{eq:app-mmin-rec}\\[4pt]
  M_{\max}(t,q)
  &= \max_{a\in \mathcal{C}_{\operatorname{op}}(t+1)}
     \bigl[ M_{\max}(t+1,q-Q(a)) + M(a) \bigr], 
  \label{eq:app-mmax-rec}
\end{align}
where terms with \(q-Q(a)<0\) or \(q-Q(a)>2L\) are forbidden and therefore treated as \(+\infty\) (for \(M_{\min}\)) and \(-\infty\) (for \(M_{\max}\)). Eqs.~\eqref{eq:app-mmin-rec} and \eqref{eq:app-mmax-rec} are evaluated for \(t=L-1,\dots,0\) using the boundary conditions at \(t=L\).

\subsubsection{Quantum-number mask at step \(t\)}

Assume a prefix \(\vek{x}_{<t}\) has already been sampled using masked logits. We define the remaining budgets with respect to the target quantum numbers \((Q^0,M^0)\) as
\begin{equation}
  q_{\mathrm{left}} = Q^0 - \sum_{i< t} Q(x_i), \qquad
  m_{\mathrm{left}} = M^0 - \sum_{i< t} M(x_i).
\end{equation}
If we choose local state \(a\in\mathcal{A}_1\) at site \(t\), the suffix \(\{t+1,\dots,L\}\) must realize
\begin{equation}
  q' = q_{\mathrm{left}} - Q(a), \qquad
  m' = m_{\mathrm{left}} - M(a).
\end{equation}
By the definitions~\eqref{eq:app-mmin-def}--\eqref{eq:app-mmax-def}, there exists at least one suffix configuration consistent with the \(\operatorname{op}\) restrictions and achieving \((q',m')\) if and only if
\begin{equation}
  0 \le q' \le 2L,\qquad
  M_{\min}(t,q') \le m' \le M_{\max}(t,q').
  \label{eq:app-qn-feasibility}
\end{equation}
We therefore define the quantum-number mask as
\begin{equation}
  m_t^{\mathrm{QN}}(a) =
  \begin{cases}
    0, & \text{if \eqref{eq:app-qn-feasibility} holds},\\[2pt]
    -\infty, & \text{otherwise.}
  \end{cases}
\end{equation}

The intuition behind this procedure is as follows. Starting from \(t=1\) with budgets \((Q^0,M^0)\), the sampling procedure at each step \(t\) draws \(x_t\) from the masked logits \(\tilde\ell_{\pmb{\theta}}=\ell_{\pmb{\theta}}+m_t^{\mathrm{op}}+m_t^{\mathrm{QN}}\) and updates \((q_{\mathrm{left}},m_{\mathrm{left}})\). By construction, \(m_t^{\mathrm{op}}\) enforces local compatibility with \(\operatorname{op}\), and \(m_t^{\mathrm{QN}}\) enforces the necessary and sufficient condition~\eqref{eq:app-qn-feasibility} for the existence of a completion.

By induction over \(t\), every prefix sampled from this procedure admits at least one completion in \(\mathcal{X}_{Q^0,M^0}\) consistent with \(\operatorname{op}\). At the final step the suffix is empty, so a completion exists if and only if the budgets are exhausted exactly, which is guaranteed by the boundary conditions on \(M_{\min/\max}\). Hence, every full configuration produced by the masked autoregressive sampler lies in \(\mathcal{X}_{Q^0,M^0}\) and respects all operator-sequence constraints.

\subsubsection{Patching procedure}

To reduce the effective sequence length of \(\vek{x}\) we group sites into non-overlapping patches of size \(\pi\) by creating larger many-body local Hilbert spaces. Assume \(\pi\) divides \(L\), then the new effective sequence length is \(P=L/\pi\). The patches are
\begin{equation}
  (1,\dots,\pi),\ (\pi{+}1,\dots,2\pi),\ \dots,\ ((P-1)\pi{+}1,\dots,P\pi),
\end{equation}
and the patch alphabet is
\begin{equation}
  \mathcal{A}_\pi = \mathcal{A}_1^{\pi},\qquad |\mathcal{A}_\pi| = 4^\pi,
\end{equation}
with the patch state \(\pmb{a} =(a_1,\dots,a_\pi)\in\mathcal{A}_\pi\). We calculate the maps \(Q\) and \(M\) in the patched case additively,
\begin{equation}
  E_\pi(\pmb{a}) = \sum_{k=1}^\pi Q(a_k),\qquad
  M_\pi(\pmb{a}) = \sum_{k=1}^\pi M(a_k).
\end{equation}
The operator sequence induces permissible patched sets
\begin{equation}
  \mathcal{C}^{(\pi)}_{\operatorname{op}}(t)
  = \Bigl\{\pmb{a}\in\mathcal{A}_\pi : a_k\in \mathcal{C}_{\operatorname{op}}((t-1)\pi+k)\ \forall k\Bigr\},
\end{equation}
for patch index \(t\in\{1,\dots,P\}\).

\subsubsection{Patched autoregressive model and QN mask}

We now express the Born distribution in terms of patches,
\begin{equation}
  p_{\pmb{\theta}}(z)
  = \prod_{t=1}^P p_{\pmb{\theta}}(z_t\mid z_{<t}),
\end{equation}
where \(z_t\in\mathcal{A}_\pi\) and \(z_{<t}\) denotes the patch prefix.
The network outputs logits \(\ell_{\pmb{\theta}}(t,\cdot\mid z_{<t})\in\mathbb{R}^{|\mathcal{A}_\pi|}\), which are masked by
\begin{equation}
  \tilde\ell_{\pmb{\theta}} = \ell_{\pmb{\theta}} + m_t^{\mathrm{op}} + m_t^{\mathrm{QN}},
\end{equation}
now defined over the patch alphabet. The patched operator mask is
\begin{equation}
  m_t^{\mathrm{op}}(\pmb{a}) =
  \begin{cases}
    0, & \pmb{a}\in \mathcal{C}^{(\pi)}_{\operatorname{op}}(t),\\[2pt]
    -\infty, & \text{otherwise.}
  \end{cases}
\end{equation}
The patched quantum-number mask \(m_t^{\mathrm{QN}}\) is constructed from patch suffix bounds \(M^{(\pi)}_{\min/\max}\), defined in analogy to Eqs.~\eqref{eq:app-mmin-def}--\eqref{eq:app-mmax-def} with \(\mathcal{A}_1\) replaced by \(\mathcal{A}_\pi\) and \((Q,M)\) by \((E_\pi,M_\pi)\). This goes as follows: for \(t\in\{0,\dots,P\}\) and \(q\in\{0,\dots,2L\}\) we define
\begin{align}
  M^{(\pi)}_{\min}(t,q)
  &= \min \{\, M_\pi(y) : y\in\mathcal{A}_\pi^{P-t},\ E_\pi(y)=q,\nonumber\\[-4pt]
  &\hspace{1.5cm}
     y \text{ obeys } \mathcal{C}^{(\pi)}_{\operatorname{op}} \,\},\\
  M^{(\pi)}_{\max}(t,q)
  &= \max \{\, M_\pi(y) : y\in\mathcal{A}_\pi^{P-t},\ E_\pi(y)=q,\nonumber\\[-4pt]
  &\hspace{1.5cm}
     y \text{ obeys } \mathcal{C}^{(\pi)}_{\operatorname{op}} \,\}.
\end{align}
The recursion is identical in structure to Eqs.~\eqref{eq:app-mmin-rec}--\eqref{eq:app-mmax-rec}:
\begin{align}
  M^{(\pi)}_{\min}(t,q)
  &= \min_{\pmb{a}\in \mathcal{C}^{(\pi)}_{\operatorname{op}}(t+1)}
     \bigl[ M^{(\pi)}_{\min}(t+1,q-E_\pi(\pmb{a})) + M_\pi(\pmb{a}) \bigr],\\[4pt]
  M^{(\pi)}_{\max}(t,q)
  &= \max_{\pmb{a}\in \mathcal{C}^{(\pi)}_{\operatorname{op}}(t+1)}
     \bigl[ M^{(\pi)}_{\max}(t+1,q-E_\pi(\pmb{a})) + M_\pi(\pmb{a}) \bigr],
\end{align}
with \(M^{(\pi)}_{\min}(P,0)=M^{(\pi)}_{\max}(P,0)=0\) and \(M^{(\pi)}_{\min}(P,q)=+\infty\), \(M^{(\pi)}_{\max}(P,q)=-\infty\) for \(q\neq 0\).

At patch step \(t\), with patch prefix \(z_{<t}\), the remaining budgets with respect to the target quantum numbers are
\begin{equation}
  q_{\mathrm{left}} = Q^0 - \sum_{u< t} E_\pi(z_u),\qquad
  m_{\mathrm{left}} = M^0 - \sum_{u< t} M_\pi(z_u).
\end{equation}
For a candidate patch \(\pmb{a}\), the suffix must realize
\begin{equation}
  q' = q_{\mathrm{left}} - E_\pi(\pmb{a}),\qquad
  m' = m_{\mathrm{left}} - M_\pi(\pmb{a}).
\end{equation}
The patched quantum-number feasibility condition is
\begin{equation}
  0 \le q' \le 2L,\qquad
  M^{(\pi)}_{\min}(t,q') \le m' \le M^{(\pi)}_{\max}(t,q'),
\end{equation}
and we set
\begin{equation}
  m_t^{\mathrm{QN}}(\pmb{a}) =
  \begin{cases}
    0, & \text{if this condition holds},\\[2pt]
    -\infty, & \text{otherwise.}
  \end{cases}
\end{equation}

Since patches are non-overlapping and the additive quantum numbers \(Q(\cdot)\) and \(M(\cdot)\) are extensive, there is a one-to-one correspondence between site-level and patched configurations that preserves \((Q(\vek{x}),M(\vek{x}))\) and operator feasibility. As a consequence, the reachable \((q,m)\) pairs on any suffix are the same whether computed sitewise or patchwise, and the patched \(m^{\mathrm{QN}}\) is equivalent to a site-level \(m^{\mathrm{QN}}\) evaluated at patch boundaries.

\section{Details of numerical renormalization group simulation}

For the single- and two-orbital impurity problems the reference data were obtained using full-density matrix NRG \cite{weichselbaum2007sum} for a Wilson-chain discretization parameter of \(\Lambda = 3\), keeping up to \(4000\) states per iteration and a maximum Wilson chain length of \(L_c = 35\). The dynamical correlators were computed at temperature \(T = 10^{-6}\).

For the three-orbital impurity problem, we constructed an effective NRG reference by attaching three additional Wilson-chain bath sites with \(\Lambda = 12\) to the third impurity orbital and then performing a standard two-channel NRG calculation with \(8000\) kept states. The Green’s function was converged with respect to the discretization parameter \(\Lambda\) and the number of kept states. Since we do not perform a dynamical mean-field self-consistency here, we only require the spectral function of a single bath leg as a reference for the NQS+SCOL results.

\end{appendix}



\bibliography{refs}


\end{document}